\documentclass[journal]{IEEEtran}
\usepackage[bookmarks,colorlinks,linkcolor=black,citecolor=black,urlcolor=black]{hyperref} % add bookmarks
\usepackage{booktabs} %\toprule、\midrule、\bottomrule
\usepackage{enumitem} % set the itemize

\usepackage{cite}
\ifCLASSINFOpdf
   \usepackage[pdftex]{graphicx}
\else
\fi
\usepackage{amssymb, mathrsfs, amsthm}
\usepackage[cmex10]{amsmath}
\begin{document}
%
% paper title
% Titles are generally capitalized except for words such as a, an, and, as,
% at, but, by, for, in, nor, of, on, or, the, to and up, which are usually
% not capitalized unless they are the first or last word of the title.
% Linebreaks \\ can be used within to get better formatting as desired.
% Do not put math or special symbols in the title.
\title{Stochastic Model Predictive Control for Linear Systems with Unbounded Additive Uncertainties}
%
%
% author names and IEEE memberships
% note positions of commas and nonbreaking spaces ( ~ ) LaTeX will not break
% a structure at a ~ so this keeps an author's name from being broken across
% two lines.
% use \thanks{} to gain access to the first footnote area
% a separate \thanks must be used for each paragraph as LaTeX2e's \thanks
% was not built to handle multiple paragraphs
%

\author{Fei~Li,
        Huiping~Li,~\IEEEmembership{Member,~IEEE},
        and~Yuyao~He,~\IEEEmembership{Member,~IEEE}% <-this % stops a space
\thanks{This work was supported in part by the National Natural Science Foundation of China (NSFC) under Grant 61922068, 61733014; in part by Shaanxi Provincial Funds for Distinguished Young Scientists under Grant 2019JC-14; in part by Aoxiang Youth Scholar Program under Grant 20GH0201111. \textit{(Corresponding author: Huiping Li)}}% <-this % stops a space

\thanks{F. Li, H. Li and Y. He are with the School of Marine Science and Technology, Northwestern Polytechnical University, Xi'an 710072, China (e-mail: lifei100400@mail.nwpu.edu.cn; lihuiping@nwpu.edu.cn; heyyao@nwpu.edu.cn).}}
% note the % following the last \IEEEmembership and also \thanks - 
% these prevent an unwanted space from occurring between the last author name
% and the end of the author line. i.e., if you had this:
% 
% \author{....lastname \thanks{...} \thanks{...} }
%                     ^------------^------------^----Do not want these spaces!
%
% a space would be appended to the last name and could cause every name on that
% line to be shifted left slightly. This is one of those "LaTeX things". For
% instance, "\textbf{A} \textbf{B}" will typeset as "A B" not "AB". To get
% "AB" then you have to do: "\textbf{A}\textbf{B}"
% \thanks is no different in this regard, so shield the last } of each \thanks
% that ends a line with a % and do not let a space in before the next \thanks.
% Spaces after \IEEEmembership other than the last one are OK (and needed) as
% you are supposed to have spaces between the names. For what it is worth,
% this is a minor point as most people would not even notice if the said evil
% space somehow managed to creep in.

% The paper headers
\markboth{}%
{LI \MakeLowercase{\textit{et al.}}: Stochastic Model Predictive Control for Linear Systems with Unbounded Additive Uncertainties}
% The only time the second header will appear is for the odd numbered pages
% after the title page when using the twoside option.
% 
% *** Note that you probably will NOT want to include the author's ***
% *** name in the headers of peer review papers.                   ***
% You can use \ifCLASSOPTIONpeerreview for conditional compilation here if
% you desire.

% If you want to put a publisher's ID mark on the page you can do it like
% this:
%\IEEEpubid{0000--0000/00\$00.00~\copyright~2015 IEEE}
% Remember, if you use this you must call \IEEEpubidadjcol in the second
% column for its text to clear the IEEEpubid mark.

% use for special paper notices
%\IEEEspecialpapernotice{(Invited Paper)}

% make the title area
\maketitle

% As a general rule, do not put math, special symbols or citations
% in the abstract or keywords.
\begin{abstract}
This paper presents two stochastic model predictive control methods for linear time-invariant systems subject to unbounded additive uncertainties. The new methods are developed by formulating the chance constraints into deterministic form, which are treated in analogy with robust constraints, by using the probabilistic reachable set. Firstly, the probabilistically resolvable time-varying tube-based stochastic model predictive control algorithm is designed by employing the time-varying probabilistic reachable sets as the tubes. Secondly, by utilizing the probabilistic positively invariant set, the probabilistically resolvable constant tube-based stochastic model predictive control algorithm is developed by employing the constantly tightened constraints in the entire prediction horizons. In addition, to enhance the feasibility of the algorithms, the soft constraints are imposed to the state initializations. The algorithm feasibility and closed-loop stability results are provided. The efficacy of the approaches are demonstrated by means of numerical simulations.
\end{abstract}

% Note that keywords are not normally used for peerreview papers.

% For peer review papers, you can put extra information on the cover
% page as needed:
% \ifCLASSOPTIONpeerreview
% \begin{center} \bfseries EDICS Category: 3-BBND \end{center}
% \fi
%
% For peerreview papers, this IEEEtran command inserts a page break and
% creates the second title. It will be ignored for other modes.
\IEEEpeerreviewmaketitle

\section{Introduction}

\IEEEPARstart{M}{odel} predictive control (MPC), also known as the receding horizon control, is inarguably the most widely implemented modern control strategy due to its conceptual simplicity and its eminent capability of handling complex dynamics and fulfilling system constraints, as well as compromising the optimal control performance and the computational load \cite{Kouvaritakis2016, Mayne2014, Huang2019, Karg2020, Dong2019}. The control performance depends, to a great extent, on the accuracy of the model. Therefore, the traditional MPC actually could not obtain the real optimal control performance, because the uncertainties, such as external disturbance and model mismatch, are widely existing. By assuming the uncertainty is bounded and considering the worst-case, the robust MPC is proposed to deal with model uncertainty \cite{Bemporad1999, Li2017Robust, Liu2018A}. Among the robust MPC, the tube-based MPC attracts huge attention due to its simply handling of constraints in the optimal control problem. In 2001, two typical tube-based MPC approaches are developed almost simultaneously. One of them uses the current measured state as the initial nominal state \cite{Chisci2001}, while the other one adopts the nominal state predicted at the first time instant \cite{Mayne2001}. Based on \cite{Mayne2001}, the David Mayne's tube-based MPC (DTMPC) \cite{Mayne2005} with flexible initialization method is presented to reduce the conservatism, where the nominal state can be searched near the actual state within an invariant set. However, all the robust MPC frameworks do not utilize the possibly existing statistical properties of the uncertainty, albeit the statistical information might be available in many cases and the worst-case scenarios unlikely occur in practice. Therefore, robust approaches may lead to overly conservatism \cite{Rawlings2017}. By employing chance constraints, the stochastic MPC (SMPC), which allows for an admissible level of constraint violation in a probabilistic sense, exploits the stochastic properties of uncertainty to achieve less conservatism in constraints satisfaction and gain better control performance \cite{Mesbah2016}. Generally, SMPC methods can be classified into the analytic approximation methods \cite{Korda2014, Farina2015, Moser2018} and the scenario-based methods \cite{Schildbach2014, Lorenzen2017a, Fleming2019}. The former ones reformulate the chance constraints into equivalent deterministic form by means of constraint tightening, while the latter ones generate a sufficient number of randomly samples to satisfy the chance constraints \cite{Farina2016}.

In the case of bounded uncertainty, the recursive feasiblility and stability have been established in several SMPC methods. For instance, in \cite{Cannon2011} the authors propose an SMPC scheme of time-varying constraint tightening along the prediction horizon based on the tubes of fix-shaped cross section and variable scaling which is computed offline according to chance violation. The tubes in \cite{Kouvaritakis2010} are constructed directly by making explicit use of the distribution of the additive disturbance. A recursively feasible SMPC scheme is proposed in \cite{Lorenzen2017b} by exerting additional constrains on the first predicted step. The work \cite{Li2020} guarantees recursive feasibility and asymptotic stability by computing the constraint tightening based on confidence region of uncertainty propagation and adopting the flexible initialization methods. When considering the unbounded stochastic uncertainty, it is typically impossible to guarantee recursive feasibility, since the uncertainty is indeed possibly (even with small probability) large enough to make the future optimization problem infeasible  \cite{Mesbah2016}. One straightforward approach for guaranteeing recursive feasibility of the SMPC problems with possibly unbounded uncertainties is to set the initial nominal state as the predicted value of the previous time instant. However, it disregards the most recent state measurement, without allowing for the feedback, which may degrade the closed-loop performance \cite{Farina2016}. To cope with this issue, an improved method of choosing the initialization between a closed-loop strategy and an open-loop one online is proposed \cite{Farina2013, Hewing2018b}. The key idea in this approach is to choose the closed-loop strategy using the measurement when the problem is feasible, and to choose the open-loop strategy when infeasible. Although this approach guarantees recursive feasibility, it requires to solve two optimisation problems at any time instant to decide the choice between the two initializing strategies. Another recent work \cite{Hewing2020} sets the initial nominal state equal to the predicted value of the first time instant, and introduces the indirect feedback via the cost function, which facilitates the recursive feasibility analysis. A special approach guarantees recursive feasibility for the system with unbounded disturbances in the case of that the chance constraint is defined as a discounted sum of violation probabilities on an infinite horizon \cite{Yan2018}.

However, regardless of the different assumptions for uncertainty, the nominal state initializations of the aforementioned works are mainly chosen to be the current measured state as in a typical tube-based MPC \cite{Chisci2001}, which may cause that the open-loop nominal state is not an accurate estimate of the actual closed-loop one, and more complex proofs of recursive feasibility and stability are required \cite{Mayne2018}. While some of the nominal state initializations are set as the predicted value of the previous time instant, which disregard the feedback of measured value, may degenerate the control performance.

In this paper, the model predictive control problem of linear time-invariant (LTI) system under unbounded additive uncertainties is studied. To solve the problem, we propose two tube-based SMPC schemes based on the framework of the DTMPC \cite{Mayne2005}, which fall in the category of analytic approximation method. The proposed tube-based SMPC schemes can be combined with the probabilistic reachable set (PRS) generated by any method to deal with the chance constraints. The novelty of the proposed method lies in that the flexible state initialization technique is used to reduce the conservatism and force the actual state stay within the relevant PRS around the nominal state with a specified probability. To the best of our knowledge, this framework has not been adopted in previous approaches of tube-based SMPC for unbounded uncertainties. The main contributions of this paper are as follows:
\begin{itemize}
	\item A new SMPC algorithm for the LTI system subject to unbounded uncertainties termed as the probabilistically resolvable time-varying tube-based SMPC (pTTSMPC) is developed. In particular, the PRS built from the uncertainty propagation is treated as the tube, by which the actual states will be restricted in the vicinity of the nominal states. The PRS is time-varying along the prediction horizon, and renders the tightened constraints change accordingly, resulting in a less conservative tightening strategy. With these techniques, the pTTSMPC algorithm is then designed based on the framework of the DTMPC \cite{Mayne2005} which can provide accurate prediction errors between the actual and nominal states, and offer an easier way to guarantee the feasibility and stability properties \cite{Mayne2018}.
%	\item By considering the probabilistic positively invariant (PPI) set and using its related tightened constraints in the entire prediction horizons, the probabilistically resolvable time-varying tube-based SMPC scheme will reduce to the special case, namely, the probabilistically resolvable constant tube-based SMPC (pCTSMPC) which has almost the same framework with DTMPC apart from the underlying probabilistic constraints.
	\item The theoretical properties of the developed algorithms are analyzed. In particular, the probabilistic resolvability, closed-loop chance constraints satisfaction and system stability are proved. Furthermore, the soft constraint is applied to the initial state which strengthens the probabilistic resolvability to stronger algorithm feasibility.
\end{itemize}  

The remainder of this paper is organized as follows. Section \ref{Preliminaries} introduces the problem to be solved and states the required definitions. Section \ref{sec pTTSMPC} presents the pTTSMPC scheme, and Section \ref{sec pCTSMPC} develops the pCTSMPC approach. In Section \ref{Simulation}, the numerical examples and comparison studies are presented and the paper is concluded in Section \ref{Conclusion}.

\textbf{Notations:} $\mathbb{R}$ represents the set of all real numbers. We denote the set of all integers which are equal or greater than $i$ by $\mathbb{N}_{i}$, and the set of all consecutive integers $\lbrace i, \cdots, j \rbrace$ by $\mathbb{N}_{i}^{j}$. ${x}_k$ denotes the value of a variable $x$ at time $k$, and $x_{k|t}$ denotes the $k$-step-ahead predicted value of $x$ at time $t$. The probability of the occurrence of event $A$ is indicated with $\textbf{Pr}(A)$. $ Q \succ 0 $ refers to a positive definite matrix, and $\Vert x \Vert_Q^2=x^T Qx $ represents the weighted norm. For a random variable $ x $, its expected value is indicated with $\textbf{E}(x)$. A random variable $ x $ of distribution $ \mathcal{Q}^x $ is denoted by $ x \sim \mathcal{Q}^x $, and a Gaussian distribution with mean $ \mu $ and variance $ W $ is represented as $ \mathcal{N} (\mu,W)$. The notation $ A \oplus B= \lbrace a+b | a \in A,b \in B \rbrace $ denotes the Minkowski sum, and $ A \ominus B = \lbrace a \in A | a+b \in A,\forall b \in B \rbrace $ represents the Pontryagin set difference.

\section{Problem Statement and Preliminaries}\label{Preliminaries}

\subsection{Problem Statement}

Consider a discrete-time LTI system subject to unbound additive uncertainties
\begin{equation}\label{system dynamics}
x_{k+1} = Ax_k + Bu_k + w_{k},
\end{equation}
where $x_k \in \mathbb R^n$, $u_k \in \mathbb R^m$ and  $w_k \in \mathbb R^n$ are the system state, control input and uncertainty at time $k$, respectively. It is assumed that all the system states are measured perfectly, and the unbounded uncertainty $ w_k \sim \mathcal{Q}^w $ is independently and identically distributed (i.i.d.). Moreover, the system in (\ref{system dynamics}) is subject to the following constraints on states and inputs
\begin{equation}\label{origin constraint}
x_k \in \mathbb{X}, u_k \in \mathbb{U}, k \in \mathbb{N}_0,
\end{equation}
where $ \mathbb{X} \subset \mathbb{R}^n $, and $ \mathbb{U} \subset \mathbb{R}^m $ are compact and each set contains the origin in its interior.

Taking advantage of the stochastic nature of the system dynamics, the chance constraints for all the predicted states and inputs in future horizon, which allow constraint violation with a probability no greater than $\varepsilon \in (0,1)$, is adopted to reduce the conservatism. The objective of this paper is to design SMPC algorithms to stabilize the system in (\ref{system dynamics}) under the following chance constraints:
\begin{subequations}
\begin{align}
\textbf{Pr}(x_{k} \in \mathbb{X}) \geq 1-\varepsilon, k \in \mathbb{N}_0 , \label{chance state constr} \\
\textbf{Pr}(u_{k} \in \mathbb{U}) \geq 1-\varepsilon, k \in \mathbb{N}_0 . \label{chance input constr}
\end{align}
\end{subequations}

\subsection{Preliminaries}

In this subsection, we recall some definitions which will facilitate algorithms design.

\newtheorem{definition}{Definition}
\begin{definition}[Robust Positively Invariant Set \cite{Rosolia2018}] \label{def 1} 
A set $ Z $ is said to be a robust positively invariant (RPI) set for the discrete-time system $ x_{k+1}=f(x_k,w_k)$, if $ x_{k+1} \in Z $, $k \in \mathbb N_0$, when $x_0 \in Z$.
\end{definition}

\begin{definition}[Probabilistic Positively Invariant Set \cite{Kofman2012}]  \label{def 2}
A set $\mathcal{D}$ is said to be a probabilistic positively invariant (PPI) set of probability level $ p $ for the discrete-time system $ x_{k+1}=f(x_k,w_k)$, if $ \textbf{Pr}(x_{k+1} \in \mathcal{D}) \geqslant p $, $k \in \mathbb N_0$, when $x_0 \in \mathcal{D}$.
\end{definition}

\begin{definition}[Confidence Region \cite{Hewing2018c}]  \label{def 3}
A set $\mathcal{E}_p$ is said to be a confidence region of probability level $ p $ for a random variable $ x $, if $ \textbf{Pr}(x \in \mathcal{E}_p) \geqslant p $.
\end{definition}

\begin{definition}[$n$-step Probabilistic Reachable Set \cite{Hewing2018b, Hewing2020}]  \label{def 4}
A set $\mathcal{R}_n$ is an $n$-step probabilistic reachable set ($n$-step PRS) of probability level $ p $ for the discrete-time system $ x_{k+1}=f(x_k,w_k)$, if $ \textbf{Pr}(x_n \in \mathcal{R}_n) \geqslant p $, initialized from $x_0$.
\end{definition}

\begin{definition}[Probabilistic Resolvability \cite{Ono2012}] \label{def 5} 
An optimization problem is probabilistically resolvable if it has feasible solutions at future time steps with a certain probability, given a feasible solution at the current time.
\end{definition}

\begin{definition}[Almost Surely Asymptotical Stability \cite{Hashimoto2013}] \label{def 6} 
A discrete-time system $ x_{k+1}=f(x_k,w_k)$ is almost surely asymptotically stable in the mean if $ \textbf{Pr} ( \lim \limits_{k \rightarrow \infty} \textbf{E}(x_k) = 0 )  = 1 $.
\end{definition}

\section{Probabilistic Time-varying Tube-based SMPC} \label{sec pTTSMPC}

\subsection{Problem Decoupling}

In order to derive a computationally tractable MPC algorithm, we make use of the techniques from DTMPC \cite{Mayne2005} to decouple the nominal trajectory planning and the uncertainty handling. Then, the system dynamics in (\ref{system dynamics}) can be separated into a nominal dynamics and an error dynamics as
\begin{subequations}
\begin{align}
& s_{k+1}=As_k+Bv_k, \label{nominal dynamics} \\
& e_{k+1}=A_{cl}e_k+w_k,  \label{error dynamics}
\end{align}
\end{subequations}
where the nominal state is denoted by $s_k \in \mathbb R^n$ and the nominal input by $v_k \in \mathbb R^m$. Without loss of generality, the feedback gain $K$ is chosen as the solution of the linear quadratic regulator (LQR) problem for the dynamics $ s_{k+1}=As_k+Bv_k $ to ensure that $ A_{cl}=A+BK $ is stable. 
The state feedback control law is used as in DTMPC
\begin{equation}\label{control policy}
u_k = Ke_k + v_k.
\end{equation}
The state error $ e_k $ between the actual state $x_k$ and the nominal state $s_k$, and the input error $ e_k^u = Ke_k $ between the control input $u_k$ and the nominal input $v_k$ are defined as
\begin{subequations}
\begin{align}
& e_k=x_k - s_k,  \label{state error} \\
& e_k^u=u_k - v_k.  \label{input error}
\end{align}
\end{subequations}

\subsection{Probabilistic Reachable Set}

Suppose that $ \mathcal E^w_{1-\varepsilon} $ is a confidence region of  probability level $ 1-\varepsilon $ for the uncertainty $ w $, where $\varepsilon \in (0,1)$. Then 
\begin{equation}\label{confidence coeff w}
\textbf{Pr}(w \in \mathcal E^w_{1-\varepsilon}) \geq 1-\varepsilon 
\end{equation}
follows from Definition \ref{def 3}. The formulation methods for confidence region of uncertainty, such as the Chebyshev inequality method \cite{Farina2015}, the Scenario generation method \cite{Farina2016}, the Box-shaped method and the Ellipsoidal method \cite{Hewing2018c}, can be incorporated with the proposed tube-based SMPC scheme.

Suppose that the confidence region $\mathcal E^{e_k}_{1-\varepsilon}$ is a $k$-step PRS of probability level $ 1-\varepsilon $ for the state error dynamics in (\ref{error dynamics}), where $\varepsilon \in (0,1)$. Then,
\begin{equation}\label{confidence coeff ex}
 \textbf{Pr}(e_k \in \mathcal E^{e_k}_{1-\varepsilon}) \geq 1-\varepsilon , k \in \mathbb N_0 , 
\end{equation}
follows from Definition \ref{def 4} directly. The $k$-step PRS of probability level $ 1-\varepsilon $ for the input error can be calculated as $ K \mathcal E^{e_k}_{1-\varepsilon} $ due to $ e_k^u = K e_k $. Then
\begin{equation}\label{confidence coeff eu}
 \textbf{Pr}(e_k^u \in K \mathcal E^{e_k}_{1-\varepsilon}) \geq 1-\varepsilon , k \in \mathbb N_0 , 
\end{equation}
is derived.

Any method to compute nested PRS, namely $ \mathcal E^{e_k}_{1-\varepsilon} \subseteq \mathcal E^{e_{k+1}}_{1-\varepsilon} $, $ k \in \mathbb N_0 $, can be combined with the proposed tube-based SMPC scheme. For instance, the Ellipsoidal method and Half-space method \cite{Hewing2020} can be adopted to produce the PRS for uncertainties with mean and variance information. For uncertainty of any type, provided that samples can be obtained, the Scenario approach \cite{Hewing2020b} can be used to formulate the PRS. And the above formulated $k$-step PRS has the property that $ \mathcal E^{e_{k+1}}_{1-\varepsilon} \subseteq A_{cl} \mathcal E^{e_k}_{1-\varepsilon} \oplus \mathcal E^w_{1-\varepsilon} $, $ k \in \mathbb N_0 $ \cite{Hewing2018c}.

Especially, consider that the uncertainty $ w $ follows the normal distribution $ w \sim \mathcal N(0,W) $. Let $ 1-\varepsilon $ be the confidence coefficient of the uncertainty, where $\varepsilon \in (0,1)$. Then the quantile value corresponding to $ 1-\varepsilon $ is $ \alpha=\Phi^{-1} (1-\varepsilon) $, where $ \Phi^{-1} $ represents the quantile function of the standard normal distribution \cite{Moser2018}. Then, the uncertainty confidence region of probability level $ 1-\varepsilon $ can be obtained as
\begin{equation}\label{confidence region w}
 \mathcal E^w_{1-\varepsilon} \triangleq  \lbrace  w: - \alpha W^{1/2} \leq w \leq \alpha W^{1/2} \rbrace.
\end{equation}
Since the error dynamics in (\ref{error dynamics}) is linear, the distribution of $ e_k $ is still normal. Then, the variance propagation of uncertainty can be evolved as
\begin{equation}\label{uncertainty variance pro}
 \Sigma_{k+1}=A_{cl} \Sigma_{k} A_{cl}^T + W, k \in \mathbb N_0,
\end{equation}
where $\Sigma_0=W$. In addition, $ \Sigma_{k+1}  \geq  \Sigma_{k} $ follows from the expanded representation of above equation.
The $k$-step PRS of probability level $ 1-\varepsilon $ can be defined as
\begin{equation}\label{confidence region e}
 \mathcal E^{e_k}_{1-\varepsilon} \triangleq  \lbrace  e: - \alpha \Sigma_k^{1/2} \leq e \leq \alpha \Sigma_k^{1/2} \rbrace,
\end{equation}
which is the error confidence region of probability level $ 1-\varepsilon $, and $ \mathcal E^{e_0}_{1-\varepsilon} = \mathcal E^w_{1-\varepsilon} $. Allowing for $ \Sigma_{k+1}  \geq  \Sigma_{k} $, the corresponding probabilistic reachable sets are nested, namely, $ \mathcal E^{e_k}_{1-\varepsilon} \subseteq \mathcal E^{e_{k+1}}_{1-\varepsilon} $. Another property of the $k$-step PRS see the following lemma.

\newtheorem{lemma}{Lemma}
\begin{lemma} \label{lemma 1}  
Consider $ w \sim \mathcal N(0,W) $. Let $ \mathcal E^w_{1-\varepsilon} $ be the uncertainty confidence region defined in (\ref{confidence region w}). Let $\mathcal E^{e_k}_{1-\varepsilon} $ be the $k$-step PRS defined in (\ref{confidence region e}). Then $ \mathcal E^{e_{k+1}}_{1-\varepsilon} \subseteq A_{cl} \mathcal E^{e_k}_{1-\varepsilon} \oplus \mathcal E^w_{1-\varepsilon} $, $ k \in \mathbb N_0 $, holds.
\end{lemma}

\begin{IEEEproof}
According to (\ref{uncertainty variance pro}) and (\ref{confidence region e}), we get that \\
 $ \mathcal E^{e_{k+1}}_{1-\varepsilon} = \lbrace  e: - \alpha \Sigma_{k+1}^{1/2} \leq e \leq \alpha \Sigma_{k+1}^{1/2} \rbrace = \\
 \lbrace  e: - \alpha (A_{cl}\Sigma_kA_{cl}^T + W)^{1/2} \leq e \leq \alpha (A_{cl}\Sigma_kA_{cl}^T + W)^{1/2} \rbrace $. \\
According to (\ref{confidence region w}) and (\ref{confidence region e}),  
$  A_{cl} \mathcal E^{e_k}_{1-\varepsilon} \oplus \mathcal E^w_{1-\varepsilon} =  \\
 \lbrace  e: - \alpha A_{cl}\Sigma_k^{1/2} - \alpha W^{1/2} \leq e \leq \alpha A_{cl}\Sigma_k^{1/2} + \alpha W^{1/2} \rbrace $ is deduced. Then, $ \mathcal E^{e_{k+1}}_{1-\varepsilon} \subseteq A_{cl} \mathcal E^{e_k}_{1-\varepsilon} \oplus \mathcal E^w_{1-\varepsilon} $ can be obtained.
\end{IEEEproof}

\subsection{Constraint Handling}

Because $ \mathcal E^{e_{k+1}}_{1-\varepsilon} \subseteq A_{cl} \mathcal E^{e_k}_{1-\varepsilon} \oplus \mathcal E^w_{1-\varepsilon} $, $ k \in \mathbb N_0 $, then $ \mathcal E^{e_{k}}_{1-\varepsilon} \subseteq \sum\limits_{i=0}^k \oplus A_{cl}^i \mathcal E^w_{1-\varepsilon} $ follows from the recursively expanding. We define the relaxed PRS as
\begin{equation}\label{confidence region dk}
 \mathcal D_k \triangleq \sum\limits_{i=0}^k \oplus A_{cl}^i \mathcal E^w_{1-\varepsilon}, k \in \mathbb N_0.
\end{equation}
Then, $ \mathcal E^{e_{k}}_{1-\varepsilon} \subseteq \mathcal D_k \subseteq \mathcal D_{k+1} $ follows directly. And
\begin{equation}\label{confidence coeff dk}
\textbf{Pr}(e_k \in \mathcal D_k) \geq 1-\varepsilon, k \in \mathbb N_0, 
\end{equation}
can be gained from (\ref{confidence coeff ex}), and 
\begin{equation}\label{confidence coeff dku}
 \textbf{Pr}(e_k^u \in K \mathcal D_k) \geq 1-\varepsilon , k \in \mathbb N_0 , 
\end{equation}
can be inferred from (\ref{confidence coeff eu}).

To guarantee the chance constraints, we propose a strategy to formulate the chance constraint sets as the time-varying deterministic constraints by using the relaxed PRS in (\ref{confidence region dk}). Construct the time-varying tightened state constraint set at step $ k $ as
\begin{equation}\label{adapt state constr}
 \tilde{\mathcal C}_{k} \triangleq \mathbb X \ominus \mathcal D_k,  k \in \mathbb N_0.
\end{equation}
Then, $ \tilde{\mathcal C}_{k+1} \subseteq \tilde{\mathcal C}_{k} $ holds owing to $ \mathcal D_k \subseteq \mathcal D_{k+1} $.
It can be seen that, if the nominal state $s_{k} \in \tilde{\mathcal C}_{k}$, then the state chance constraint in (\ref{chance state constr}) is satisfied, i.e., $ \textbf{Pr}(x_{k} = s_{k} + e_{k} \in \mathbb{X}) \geq 1-\varepsilon $ can be guaranteed by (\ref{confidence coeff dk}).

The satisfaction of input chance constraint in (\ref{chance input constr}), i.e., $ \textbf{Pr}(u_{k} = v_{k} + e_{k}^u \in \mathbb{U}) \geq 1-\varepsilon $ can be guaranteed by constructing the  tightened input constraint set at step $ k $ as 
\begin{equation}\label{adapt input constr}
 \tilde{\mathcal V}_{k} \triangleq \mathbb U \ominus K \mathcal D_k,  k \in \mathbb N_0.
\end{equation}
Similarly, if the nominal input $ v_k \in \tilde{\mathcal V}_{k} $, then the input chance constraint in (\ref{chance input constr}) is satisfied according to (\ref{confidence coeff dku}).

Furthermore, let the minimal robust positively invariant (mRPI) set
\begin{equation}\label{mRPI}
 Z \triangleq \sum\limits_{i=0}^\infty \oplus A_{cl}^i \mathcal E^w_{1-\varepsilon}
\end{equation}
for the error dynamics in (\ref{error dynamics}) with $ w \in \mathcal E^w_{1-\varepsilon} $ be computed as the outer approximation using the strategy in \cite{Rakovic2005}. Then, $ \mathcal D_k \subseteq Z $ and $ K \mathcal D_k \subseteq KZ $ can be derived from (\ref{confidence region dk}).

\subsection{Formulation of pTTSMPC}

For the cost function, we adopt the stage cost $ \ell(s_k,v_k)=\Vert s_k \Vert_Q^2+\Vert v_k \Vert_R^2$, and the terminal cost $V_f=\Vert s \Vert_P^2$, where $Q \succ 0 $, $R \succ 0$, and $P$ is the solution of the algebraic Lyapunov equation
\begin{equation*}
(A+BK)^TP(A+BK)-P = -Q-K^TRK.
\end{equation*}

To ensure the feasibility, we construct the terminal constraint set as the maximal output admissible set \cite{Kolmanovsky1995}
\begin{equation}\label{terminal X_f v}
 \tilde{\mathcal X}_f \triangleq \lbrace s \in \mathbb R^n: s_k \in \tilde{\mathcal C}_{k},Ks_k \in \tilde{\mathcal V}_{k},  k \in \mathbb N_0 \rbrace.
\end{equation}
In this case, $\tilde{\mathcal X}_f$ satisfies the axioms \cite{Mayne2000}:
\begin{itemize}
\item[\textbf{A1:}]  $ A_{cl} \tilde{\mathcal X}_f \subset \tilde{\mathcal X}_f, \tilde{\mathcal X}_f \subset \mathbb X \ominus Z, K\tilde{\mathcal X}_f \subset \mathbb U \ominus KZ$.
\item[\textbf{A2:}]  $V_f (A_{cl} x)+\ell(x,Kx) \leq V_f (x), \forall x\in \tilde{\mathcal X}_f$.
\end{itemize}

On the basis of the time-varying constraint tightening method and the DTMPC framework in \cite{Mayne2005}, the probabilistic time-varying tube-based stochastic finite horizon optimal control problem $\mathbb P^{ttsmpc}_N(x_t)$ to be solved at each time instant $t$ is formulated as follows:
\begin{subequations}\label{tTSMPC}
\begin{align}
  \min_{s_{0|t}, \textbf{\textit{v}}_{t}} & \sum \limits_{k=0}^{N-1}(\Vert s_{k|t}\Vert_Q^2+\Vert v_{k|t}\Vert_R^2) + \Vert s_{N|t}\Vert_P^2 \label{cost fun v} \\
 s.t. \quad  & s_{k|t}=As_{k|t}+Bv_{k|t}, \\
             & s_{k|t} \in \tilde{\mathcal C}_{k+t},  k \in \mathbb{N}_1^{N-1} ,\label{constraints state v}  \\
             & v_{k|t} \in  \tilde{\mathcal V}_{k+t},  k \in \mathbb{N}_0^{N-1} , \label{constraints input v}   \\
             & x_t - s_{0|t} \in \mathcal D_t, \label{constraints init v}  \\
             & s_{N|t} \in \tilde{\mathcal X}_f. \label{constraints terminal v} 
\end{align}
\end{subequations}

The optimal solution of $\mathbb P^{ttsmpc}_N(x_t)$ consists of nominal state $s_{0|t}^*$ and input sequence $ \textbf{\textit{v}}^* (x_t)=[v_{0|t}^*, v_{1|t}^*, \cdots, v_{N-1|t}^* ] $, and the associated optimal state sequence for the nominal system is $ \textbf{\textit{s}}^* (x_t)=[s_{0|t}^*, s_{1|t}^*, \cdots, s_{N|t}^* ] $.

At each time instant, $\mathbb P^{ttsmpc}_N(x_t)$ is solved to generate $ \textbf{\textit{v}}^* (x_t) $ and $s_{0|t}^*$, and the  optimal control law is designed as:
\begin{equation}\label{optimal control law}
 u^*(x_t) = K(x_t-s_{0|t}^*) + v_{0|t}^*.
\end{equation}

The entire process of the probabilistic time-varying tube-based SMPC is repeated for all $t \geqslant 0$ to yield a receding horizon control strategy.

\subsection{Properties of pTTSMPC}

In this section, the feasibility of the designed pTTSMPC scheme is first presented, then the closed-loop chance constraint satisfaction is verified, and the stability results of the closed-loop system are also provided. Before that, a lemma for developing the feasibility result is presented.

\begin{lemma} \label{lemma 2}  
Let $\mathcal E^{e_t}_{1-\varepsilon}$ be the $t$-step PRS of probability level $ 1-\varepsilon $ for the state error dynamics in (\ref{error dynamics}) with $ w_t \sim \mathcal Q^w $. For $ x_{t+1} = Ax_t + Bu_t + w_t, $ and $ s_{k+1|t}=As_{k|t}+Bv_{k|t} $, if $ x_t \in s_{0|t} \oplus \mathcal D_t $ and $ u_t = K(x_t-s_{0|t}) + v_{0|t} $, then $ \textbf{Pr}(x_{t+1} \in s_{1|t} \oplus \mathcal D_{t+1}) \geqslant 1-\varepsilon $ for all $ w_t \sim \mathcal Q^w $. 
\end{lemma}

\begin{IEEEproof}
The proof follows the similar line of reasoning as Proposition 1 in \cite{Mayne2001}. In (\ref{state error}) we have defined $ e_t=x_t-s_{0|t} $ and $ e_{t+1}=x_{t+1}-s_{1|t} $. Due to $ x_t \in s_{0|t} \oplus \mathcal D_t $, it follows that $ e_t=x_t-s_{0|t} \in \mathcal D_t $. Since, according to (\ref{error dynamics}), (\ref{confidence coeff w}) and (\ref{confidence region dk}), $ \textbf{Pr}(e_{t+1}=A_{cl} e_t + w_t \in A_{cl} \mathcal D_t \oplus \mathcal E^{w}_{1-\varepsilon}) \geqslant 1-\varepsilon $ and $ \mathcal D_{t+1} = A_{cl} \mathcal D_t \oplus \mathcal E^{w}_{1-\varepsilon} $ hold, then $ \textbf{Pr}(x_{t+1}= s_{1|t} + e_{t+1} \in s_{1|t} \oplus \mathcal D_{t+1}) \geqslant 1-\varepsilon $ follows. 
\end{IEEEproof}

Based on Lemma \ref{lemma 2}, the probabilistic resolvability result is reported as follows.

\newtheorem{theorem}{Theorem}{}
\begin{theorem}[Probabilistic Resolvability]
Consider the optimal control problem $\mathbb P_{N}^{ttsmpc}(x_t)$ in (\ref{tTSMPC}) for the system dynamics in (\ref{system dynamics}) under the control law in (\ref{optimal control law}). If $ \mathcal M_t=[s_{0|t}^*,v_{0|t}^*,v_{1|t}^*,\cdots,v_{N-1|t}^* ]$ is feasible for the problem at time instant $ t $, then applying the control law in (\ref{optimal control law}) gives the probabilistic resolvability at time instant $ t+1 $.
\end{theorem}

\begin{IEEEproof}
The proof follows the similar line of reasoning as Proposition 3 in \cite{Mayne2005}. Since $ \mathcal M_t $ is feasible for the problem $\mathbb P_{N}^{ttsmpc}(x_t)$, the constraints (\ref{constraints state v})-(\ref{constraints terminal v}) are satisfied by $ \textbf{\textit{v}}^*(x_t) $ and $ \textbf{\textit{s}}^*(x_t) $. The shifted input sequence of $ \textbf{\textit{v}}^*(x_t) $ is denoted as $ \vec{\textbf{\textit{v}}}^*(x_t) =[v_{1|t}^*,\cdots,v_{N-1|t}^*,Ks_{N|t}^* ] $ and the shifted state sequence of $ \textbf{\textit{s}}^*(x_t) $ is denoted as $ \vec{\textbf{\textit{s}}}^*(x_t) =[s_{1|t}^*,\cdots,s_{N|t}^*,A_{cl} s_{N|t}^* ] $. We choose the candidate solution $ \mathcal M_{t+1}=[s_{1|t}^*,v_{1|t}^*,\cdots,v_{N-1|t}^*, Ks_{N|t}^* ]$. Hence, the first $ N $ elements of $ \vec{\textbf{\textit{s}}}^*(x_t) $ satisfy the time-varying tightened state constraints (\ref{constraints state v}) owing to the fact that if $ s_{k+1|t}^* \in \tilde{\mathcal C}_{k+1+t} $, then $ s_{k+1|t}^* \in \tilde{\mathcal C}_{k+t} $ as $ \tilde{\mathcal C}_{k+1+t} \subseteq \tilde{\mathcal C}_{k+t} $. And the first $ N-1 $ elements of $ \vec{\textbf{\textit{v}}}^*(x_t) $ satisfy the time-varying tightened input constraint (\ref{constraints input v}) analogously. Because $ s_{N|t}^* \in \tilde X_f $, it follows from $ \textbf{A1} $ that $ A_{cl} s_{N|t}^* \in \tilde X_f$ and $ Ks_{N|t}^* \in \mathbb U \ominus KZ $. So the last element $ A_{cl} s_{N|t}^* $ of $ \vec{\textbf{\textit{s}}}^*(x_t) $ satisfies the terminal constraint (\ref{constraints terminal v}). Allowing for $ K \mathcal D_{t+N} \subseteq KZ $, thus $ Ks_{N|t}^* \in \mathbb U \ominus KZ \subseteq \mathbb U \ominus \mathcal D_{t+N} $ follows. Therefore, the last element $ Ks_{N|t}^* $ of $ \vec{\textbf{\textit{v}}}^*(x_t) $ satisfies constraint (\ref{constraints input v}). Moreover, from Lemma \ref{lemma 2} we have that if $ x_t \in s_{0|t}^* \oplus \mathcal D_t $, then $ \textbf{Pr}(x_{t+1} \in s_{1|t}^* \oplus \mathcal D_{t+1}) \geqslant 1-\varepsilon $ for all  $ w_t \sim \mathcal{Q}^w $, i.e., the initial constraint (\ref{constraints init v}) is satisfied with the probability no less than $ 1-\varepsilon $. Summarizing the satisfaction status of the state constraint (\ref{constraints state v}), the input constraint (\ref{constraints input v}), the initial constraint (\ref{constraints init v}), and the terminal constraint (\ref{constraints terminal v}), $ \mathcal M_{t+1} $ is probabilistically resolvable for the problem $\mathbb P_{N}^{ttsmpc}(x_{t+1})$.
\end{IEEEproof}

Suppose that the problem $\mathbb P_{N}^{ttsmpc}(x_t)$ is feasible, then the satisfaction of closed-loop chance constraints follows directly, which is shown in the Theorem 2.

\begin{theorem}[Chance Constraint Satisfaction]
Consider the optimal control problem $\mathbb P_{N}^{ttsmpc}(x_t)$ in (\ref{tTSMPC}) for the system dynamics in (\ref{system dynamics}) under the control law in (\ref{optimal control law}). Suppose $\mathbb P_{N}^{ttsmpc}(x_t)$ is recursively feasible, then the closed-loop state chance constraint in (\ref{chance state constr}) and input chance constraint in (\ref{chance input constr}) are satisfied.
\end{theorem}

\begin{IEEEproof}
From the feasibility of the optimal control problem $\mathbb P_{N}^{ttsmpc}(x_t)$ in (\ref{tTSMPC}) we have that $ e_t=x_t-s_{0|t}^* \in \mathcal D_t $, $ t \in \mathbb{N}_0 $. Using the fact $ \mathcal D_{t+1} = A_{cl} \mathcal D_t \oplus \mathcal E^w_{1-\varepsilon} $, we can get $ \textbf{Pr}(A_{cl} e_t + w_t \in \mathcal D_{t+1} | e_t \in \mathcal D_t) \geq 1-\varepsilon $ for all $ w_t \sim \mathcal Q^w $ following from (\ref{confidence coeff w}). Since $ s_{1|t}^* \in \tilde{\mathcal C}_{t+1} $, $ \textbf{Pr}(s_{1|t}^* + A_{cl} e_t + w_t \in \tilde{\mathcal C}_{t+1} \oplus \mathcal D_{t+1} | e_t \in \mathcal D_t) \geq 1-\varepsilon $ holds. Consequently, as $ x_{t+1}=s_{1|t}^* + A_{cl} e_t + w_t $ and $ \mathbb X = \tilde{\mathcal C}_{t+1} \oplus \mathcal D_{t+1} $, it follows that $ \textbf{Pr}(x_{t+1} \in \mathbb X) \geq 1-\varepsilon $, $ t \in \mathbb{N}_0 $. The same argument holds for the input constraint, namely, $ \textbf{Pr}(u_{t+1} \in \mathbb U) \geq 1-\varepsilon $, $ t \in \mathbb{N}_0 $.
\end{IEEEproof}

Finally, the stability result is presented in the following theorem.

\begin{theorem}[Stability]
Consider the optimal control problem $\mathbb P_{N}^{ttsmpc}(x_t)$ in (\ref{tTSMPC}) for the system dynamics in (\ref{system dynamics}) under the control law in (\ref{optimal control law}). Suppose that $ Z $ is the mRPI set for the error dynamics in (\ref{error dynamics}) and $\mathbb P_{N}^{ttsmpc}(x_t)$ is recursively feasible, the following hold.
\begin{itemize}
\item[(a)] $ Z $ is probabilistically stable for the closed-loop system.
\item[(b)] If the uncertainty has zero mean, the closed-loop system is almost surely asymptotically stable in the mean.
\end{itemize}
\end{theorem}

\begin{IEEEproof}
Denote the cost function as $ J(\textbf{\textit{s}}_t, \textbf{\textit{v}}_t) = \sum \limits_{k=0}^{N-1}(\Vert s_{k|t}\Vert_Q^2+\Vert v_{k|t}\Vert_R^2) + \Vert s_{N|t}\Vert_P^2 $, and the optimal cost value as $J^*(\textbf{\textit{s}}_t, \textbf{\textit{v}}_t) = \min J(\textbf{\textit{s}}_t, \textbf{\textit{v}}_t)$. From the Proposition 3 in \cite{Mayne2005} we have that $ J^*(\textbf{\textit{s}}_{t+1}, \textbf{\textit{v}}_{t+1}) - J^*(\textbf{\textit{s}}_t, \textbf{\textit{v}}_t) \leq - \Vert s_{0|t}\Vert_Q^2 - \Vert v_{0|t}\Vert_R^2 $. Summing both sides of above inequality over all $ t \geq 0 $ gives the closed-loop performance bound as $ \sum\limits_{t=0}^\infty (\Vert s_{0|t}\Vert_Q^2+\Vert v_{0|t}\Vert_R^2) \leq J^*(\textbf{\textit{s}}_{0}, \textbf{\textit{v}}_{0}) - J^*(\textbf{\textit{s}}_\infty, \textbf{\textit{v}}_\infty) $. The left hand side of the inequality is the cost value along the nominal trajectory. Since $ J^*(\textbf{\textit{s}}_{t}, \textbf{\textit{v}}_{t}) \geq 0 $ for all $ t $, it follows that $ \sum\limits_{t=0}^\infty (\Vert s_{0|t}\Vert_Q^2+\Vert v_{0|t}\Vert_R^2) \leq J^*(\textbf{\textit{s}}_{0}, \textbf{\textit{v}}_{0}) $. Given that the optimal cost is necessarily finite if the problem $\mathbb P_{N}^{ttsmpc}(x_t)$ is feasible, and since each term on the left hand side of the above inequality is non-negative, $ \lim \limits_{t \rightarrow \infty} (\Vert s_{0|t}\Vert_Q^2+\Vert v_{0|t}\Vert_R^2) = 0 $ follows. Furthermore, since $Q \succ 0 $, $R \succ 0$, it implies that $ \lim \limits_{k \rightarrow \infty} s_{k|t} =0 $ following from Theorem 2.7 in \cite{Kouvaritakis2016}. Consequently, the nominal origin $ s=0 $ is exponentially stable for the nominal dynamics by following from Theorem 2.8 in \cite{Kouvaritakis2016}. Moreover, $ \textbf{Pr}(e_{t+k} \in Z) \geqslant 1-\varepsilon $ holds, due to $ \textbf{Pr}(e_{t+k} \in \mathcal D_{t+k}) \geqslant 1-\varepsilon $ and $ \mathcal D_{t+k} \subseteq Z $, hence $ \textbf{Pr} ( \lim \limits_{k \rightarrow \infty} x_{t+k} = \lim \limits_{k \rightarrow \infty} ( s_{k|t}+e_{t+k} ) = \lim \limits_{k \rightarrow \infty} e_{t+k} \in Z )  \geqslant 1-\varepsilon $. This completes the proof for part (a).

The proof of part (b) is as follows. Since the uncertainty is zero-mean, allowing for the error dynamics in (\ref{error dynamics}), $ \textbf{Pr} ( \lim \limits_{k \rightarrow \infty} \textbf{E}(e_{t+k}) = 0 )  = 1 $ is derived directly. Then $ \textbf{Pr} ( \lim \limits_{k \rightarrow \infty} \textbf{E}(x_{t+k})=0) = \textbf{Pr} ( \lim \limits_{k \rightarrow \infty} \textbf{E} ( s_{k|t}+e_{t+k} ) = 0)= \textbf{Pr} ( \lim \limits_{k \rightarrow \infty} \textbf{E}(e_{t+k}) = 0 ) = 1 $. Thus, according to Definition \ref{def 6}, the closed-loop system is almost surely asymptotically stable in the mean.
\end{IEEEproof}

\subsection{Enhanced Feasibility}

In order to avoid possible infeasibility when violation occurs, the slack variable can be imposed on the state constraints \cite{Korda2011, Moser2018, Paulson2020}. However, here, we only impose the slack variable $ \lambda $ on the flexible initial state constraint in (\ref{constraints init v}), such that
\begin{equation}\label{soft init}
  x_t - s_{0|t} \in \lambda  \mathcal D_t ,  \lambda  \geq  1. 
\end{equation}
And a finite penalty term  
\begin{equation}\label{penalty cost}
\psi(\lambda) = \gamma ( \frac{1}{1 + e^{- (\lambda - 1) }} - \frac{1}{2} ) 
\end{equation}
is added to the cost function in (\ref{cost fun v}) to force $ \lambda $ to tend towards 1, so that the distance between the actual state and the nominal state will be as close as possible while maintaining the feasibility. The weight $ \gamma $ is chosen large enough so that $ \psi(\lambda) $ makes the best of the penalty function \cite{Carpintero2018}. As a result, the probabilistic resolvability of the optimal control problem $\mathbb P^{ttsmpc}_N(x_t)$ is enhanced to strong feasibility by using the soft initial state constraint in (\ref{soft init}) and the penalty term in (\ref{penalty cost}).
Moreover, this approach allows for solving a single optimisation problem instead of having to verify feasibility and decide which MPC problem to be solved accordingly, as is the case in \cite{Farina2013, Farina2015, Hewing2018b}.

\section{Probabilistic Constant Tube-based SMPC} \label{sec pCTSMPC}

\subsection{Probabilistic Positively Invariant Set}

The relationship between the RPI set and the PPI set is presented in the following proposition.

\newtheorem{proposition}{Proposition}
\begin{proposition} \cite{Hewing2018c}\label{propos 1}  
Let $ Z $ be a RPI set for the error dynamics in (\ref{error dynamics}) with $ w \in \mathcal E^w_{1-\varepsilon} $, where $ \mathcal E^w_{1-\varepsilon} $ is the confidence region of probability $ 1-\varepsilon $ for $ w $. If  $ \mathcal E^{e_{k+1}}_{1-\varepsilon} \subseteq A_{cl} \mathcal E^{e_k}_{1-\varepsilon} \oplus \mathcal E^w_{1-\varepsilon} $, $ k \in \mathbb N_0 $, then $ Z $ is also a PPI set of probability level $ 1-\varepsilon $ for the error dynamics in (\ref{error dynamics}) with $ w \sim \mathcal Q^w $.
\end{proposition}

Let $ Z $ be the mRPI set for the error dynamics in (\ref{error dynamics}) with $ w \in \mathcal E^w_{1-\varepsilon} $ computed as in (\ref{mRPI}). If $ \mathcal E^w_{1-\varepsilon} $ is a box-shaped or an ellipsoidal confidence region, then $ Z $ is also the minimal probabilistic positively invariant (mPPI) set of probability level $ 1-\varepsilon $ for the error dynamics in (\ref{error dynamics}) with $ w \sim \mathcal Q^w $ \cite{Hewing2018c}.

Especially, when the uncertainty $ w $ follows the normal distribution  $ w \sim \mathcal N(0,W) $, the following property holds.

\newtheorem{corollary}{Corollary}
\begin{corollary} \label{corol 1}  
Let $ Z $ be the mRPI set for the error dynamics in (\ref{error dynamics}) with $ w \in \mathcal E^w_{1-\varepsilon} $, where $ \mathcal E^w_{1-\varepsilon} $ is defined in (\ref{confidence region w}). Then $ Z $ is also the mPPI set of probability level $ 1-\varepsilon $ for the error dynamics in (\ref{error dynamics}) with $ w \sim \mathcal N(0,W) $.
\end{corollary}

\begin{IEEEproof}
According to Lemma \ref{lemma 1}, $ \mathcal E^{e_{k+1}}_{1-\varepsilon} \subseteq A_{cl} \mathcal E^{e_k}_{1-\varepsilon} \oplus \mathcal E^w_{1-\varepsilon} $, $ k \in \mathbb N_0 $. Thus, the claim follows from the Proposition \ref{propos 1}.
\end{IEEEproof}

\subsection{Formulation of pCTSMPC}

%Consider $Z$ is the mPPI set of probability level $ 1-\varepsilon $ for the error dynamics in (\ref{error dynamics}) with $ w \sim \mathcal Q^w $, $\mathcal E^{e_k}_{1-\varepsilon}$ is the $k$-step PRS of probability level $ 1-\varepsilon $ for the error dynamics in (\ref{error dynamics}), and $ \mathcal E^w_{1-\varepsilon} $ is the confidence region of probability $ 1-\varepsilon $ for $ w $. Note that, since $ \mathcal E^{e_{k}}_{1-\varepsilon} \subseteq \sum\limits_{i=0}^k \oplus A_{cl}^i \mathcal E^w_{1-\varepsilon} $, then $ \lim \limits_{t \rightarrow \infty} \mathcal E^{e_{t}}_{1-\varepsilon}  \subseteq  \sum\limits_{i=0}^{\infty} \oplus A_{cl}^i \mathcal E^w_{1-\varepsilon} = Z $, which is defined in (\ref{mRPI}), namely $ \mathcal E^{e_\infty}_{1-\varepsilon}  \subseteq  Z $. As a result, $ \mathcal E^{e_k}_{1-\varepsilon}  \subseteq  Z $, $ k \in \mathbb N_0 $, follows from $ \mathcal E^{e_k}_{1-\varepsilon} \subseteq \mathcal E^{e_{k+1}}_{1-\varepsilon} $. Consequently,

Since $Z$ is the mPPI set of probability level $ 1-\varepsilon $ for the error dynamics in (\ref{error dynamics}) with $ w \sim \mathcal Q^w $, then
\begin{equation}\label{confidence coeff ex c}
 \textbf{Pr}(e_k \in Z) \geq 1-\varepsilon , k \in \mathbb N_0 , 
\end{equation}
follows from Definition \ref{def 2} directly. And
\begin{equation}\label{confidence coeff eu c}
 \textbf{Pr}(e_k^u \in K Z) \geq 1-\varepsilon , k \in \mathbb N_0 , 
\end{equation}
can be derived from $ e_k^u = K e_k $ subsequently.

Construct the constantly tightened state constraint set as
\begin{equation}\label{tightened state constr}
\bar{\mathcal C} \triangleq \mathbb X \ominus Z,
\end{equation}
then, the chance constraint in (\ref{chance state constr}) is satisfied. The reason is that, if the nominal state $s_{k} \in \bar{\mathcal C}$, then, $ \textbf{Pr}(x_{k} = s_{k} + e_{k} \in \mathbb{X}) \geq 1-\varepsilon $ can be guaranteed by (\ref{confidence coeff  ex c}).

The satisfaction of $ \textbf{Pr}(u_{k} = v_{k} + e_{k}^u \in \mathbb{U}) \geq 1-\varepsilon $ can be guaranteed by constructing the constantly tightened input constraint set as 
\begin{equation}\label{tightened input constr}
 \bar{\mathcal V} \triangleq \mathbb U \ominus K Z,  k \in \mathbb N_0.
\end{equation}
Thus, if the nominal input $ v_k \in \bar{\mathcal V} $, then the input chance constraint in (\ref{chance input constr}) is satisfied according to (\ref{confidence coeff eu c}).

Furthermore, to ensure the feasibility, we construct the terminal constraint set as
\begin{equation}\label{terminal X_f c}
\bar X_f \triangleq \lbrace s \in \mathbb R^n: s_k \in \bar{\mathcal C},Ks_k \in \bar{\mathcal V},  k \in \mathbb N_0  \rbrace,
\end{equation}
which satisfies the axioms \textbf{A1} and \textbf{A2} similar with (\ref{terminal X_f v}).

On the basis of the constant constraint tightening method and the DTMPC framework in \cite{Mayne2005}, the probabilistic constant tube-based stochastic finite horizon optimal control problem $\mathbb P^{ctsmpc}_N(x_t)$ to be solved at each time instant $t$ is formulated as follows:
\begin{subequations}\label{cTSMPC}
\begin{align}
  \min_{s_{0|t}, \textbf{\textit{v}}_{t}} & \sum \limits_{k=0}^{N-1}(\Vert s_{k|t}\Vert_Q^2+\Vert v_{k|t}\Vert_R^2) + \Vert s_{N|t}\Vert_P^2  \label{cost fun c} \\
 s.t. \quad  & s_{k|t}=As_{k|t}+Bv_{k|t}, \\
             & s_{k|t} \in \bar{\mathcal C},  k \in \mathbb{N}_1^{N-1} ,  \\
             & v_{k|t} \in \bar{\mathcal V},  k \in \mathbb{N}_0^{N-1} ,   \\
             & x_t - s_{0|t} \in Z ,  \label{constraints init c}  \\
             & s_{N|t} \in \bar{\mathcal X}_f.  
\end{align}
\end{subequations}

The optimal control law is designed as in (\ref{optimal control law}). At each time instant, $\mathbb P^{ctsmpc}_N(x_t)$ is solved to generate  input sequence $ \textbf{\textit{v}}^* (x_t) $ and nominal state $s_{0|t}^*$. The entire process of the probabilistic constant tube-based SMPC is repeated for all $t \geqslant 0$ to yield a receding horizon control strategy.

Using the similar argument as in pTTSMPC, we can prove that the optimal control problem $\mathbb P^{ctsmpc}_N(x_t)$ in (\ref{cTSMPC}) is probabilistically resolvable with the probability no less than $ 1-\varepsilon $. Suppose that the problem $\mathbb P^{ctsmpc}_N(x_t)$ is recursively feasible, then the satisfaction of closed-loop chance constraints can be guaranteed, and $ Z $ is probabilistically stable for the closed-loop system. Moreover, if the uncertainty has zero mean, the closed-loop system is almost surely asymptotically stable in the mean.

Besides, in order to avoid possible infeasibility when violation occurs, we impose the slack variable $ \lambda $ on the flexible initial state constraint in (\ref{constraints init c}), such that
\begin{equation}\label{soft init c}
  x_t - s_{0|t} \in \lambda Z ,  \lambda  \geq  1. 
\end{equation}
And a finite penalty term as in (\ref{penalty cost}) is added to the cost function in (\ref{cost fun c}) to force $ \lambda $ to tend towards 1. As a result, the probabilistic resolvability of the optimal control problem $\mathbb P^{ctsmpc}_N(x_t)$ is enhanced to strong feasibility by using the soft initial state constraint in (\ref{soft init c}).

\newtheorem{remark}{Remark}{}
\begin{remark}
In both the pTTSMPC and the pCTSMPC schemes, all the constraint sets can be calculated offline, therefore, the resulting SMPC algorithms require the comparable online computational load with the DTMPC algorithm.
\end{remark}

\section{Numerical Simulation} \label{Simulation}

In this section, the control effects of the proposed probabilistic tube-based SMPC schemes are first illustrated. Then, the chance constraint violation of the proposed methods are compared. In the simulations, the feasible regions in the figures are caculated by using the algorithm in \cite{Blanchini2008}, the implementations of the algorithms and the computations of $ \tilde{\mathcal C}_{k} $, $ \tilde{\mathcal V}_{k} $, $ \bar{\mathcal C} $, $ \bar{\mathcal V} $ and $ Z $ are performed by using the MPT3 toolbox \cite{Herceg2013}.

Consider a discrete-time LTI system in \cite{Rosolia2018} subject to additive uncertainties with unbounded support
\begin{align}\label{example_system_1}
		x_{k+1}=
	\begin{bmatrix}
		1&0.0075\\-0.143&0.996
	\end{bmatrix}x_k+
	\begin{bmatrix}
		4.798\\0.115
	\end{bmatrix}u_k+w_k,
\end{align}
where $ \mathbb{X}\triangleq \left\lbrace  x \in \mathbb R^2: \begin{bmatrix} -2\\ -3 \end{bmatrix} \leq x \leq \begin{bmatrix} 2\\3 \end{bmatrix} \right\rbrace  $, $ \mathbb{U}\triangleq \lbrace u \in \mathbb R: \vert u \vert \leq 0.2 \rbrace $, and $w_k \sim \mathcal N(w_k,0.04^2)$. The state and input constraints are $ \textbf{Pr}(x_{k} \in \mathbb{X}) \geq 0.8 $ and $ \textbf{Pr}(u_{k} \in \mathbb{U}) \geq 0.8 $, $ k \in \mathbb N_0 $. The weights of the cost functions are $ Q=\begin{bmatrix} 1&0\\0&10 \end{bmatrix} $ and $ R=1 $, respectively. $ K $ is the LQR feedback gain for the unconstrained optimal problem $ (A,B,Q,R) $. The prediction horizon is $ N=8 $, the simulation step is $ N_{\rm sim}=15 $ and the initial state $ x_0=[2.5,2.8]^T $. Set the weight $ \gamma = 100 $ in the cost function of the strongly feasible SMPC approaches.

The simulation of the pTTSMPC, pTTSMPC with enhanced feasibility (denoted as pTTSMPC-en), pCTSMPC and pCTSMPC with enhanced feasibility (denoted as pCTSMPC-en) for the system in (\ref{example_system_1}) are conducted. The results are as follows.

\subsection{Control Effects}

\begin{figure}[tbhp]
\begin{minipage}{1\linewidth}
 \centering
  \includegraphics[width=0.95\textwidth,trim= 0 0 0 0,clip]{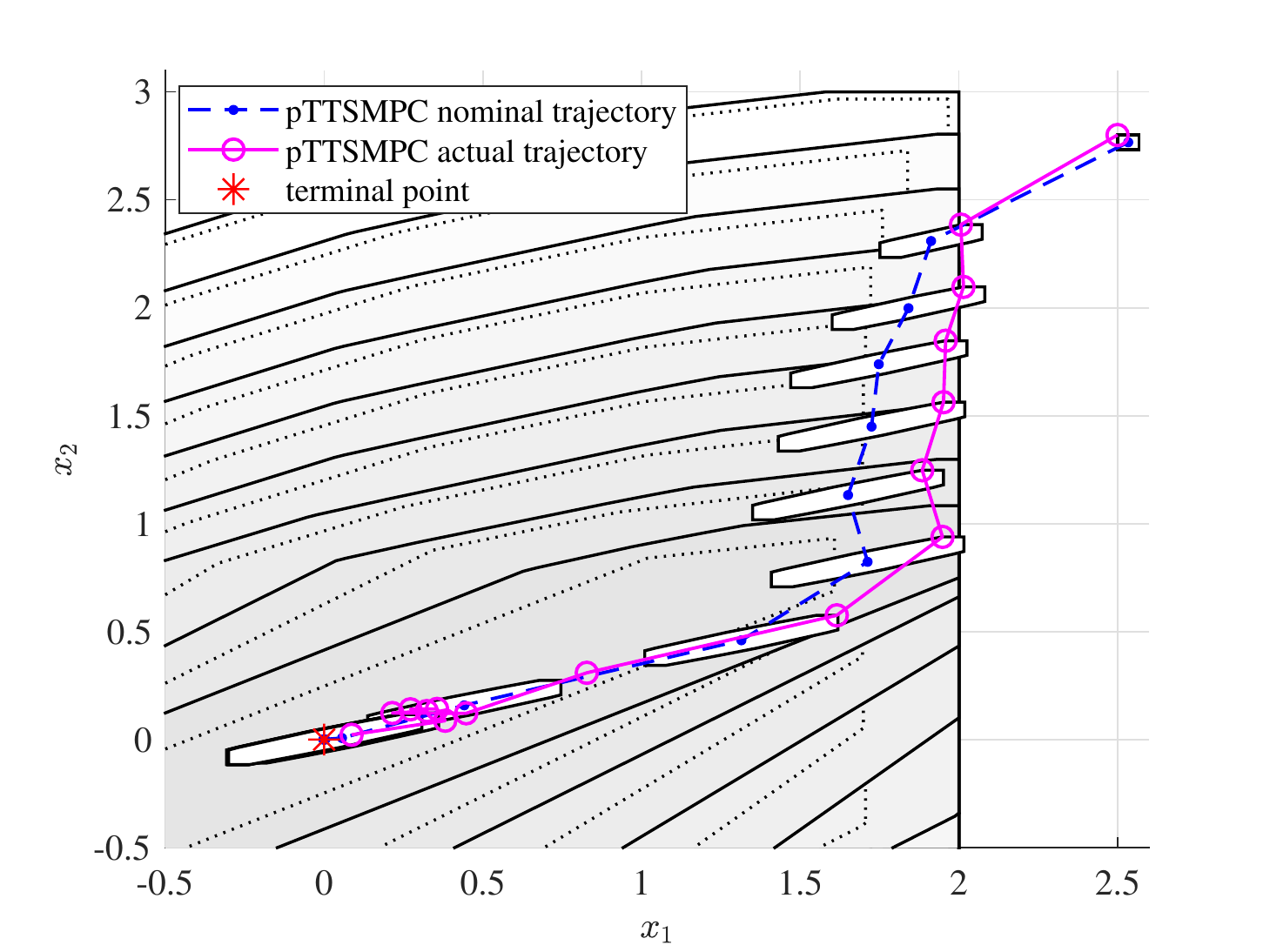}
\end{minipage}
  \caption{Closed-loop state trajectories of the pTTSMPC.}\label{Fig trajectory Vp}
\end{figure}

\begin{figure}[tbhp]
\begin{minipage}{1\linewidth}
 \centering
  \includegraphics[width=0.95\textwidth,trim= 0 0 0 0,clip]{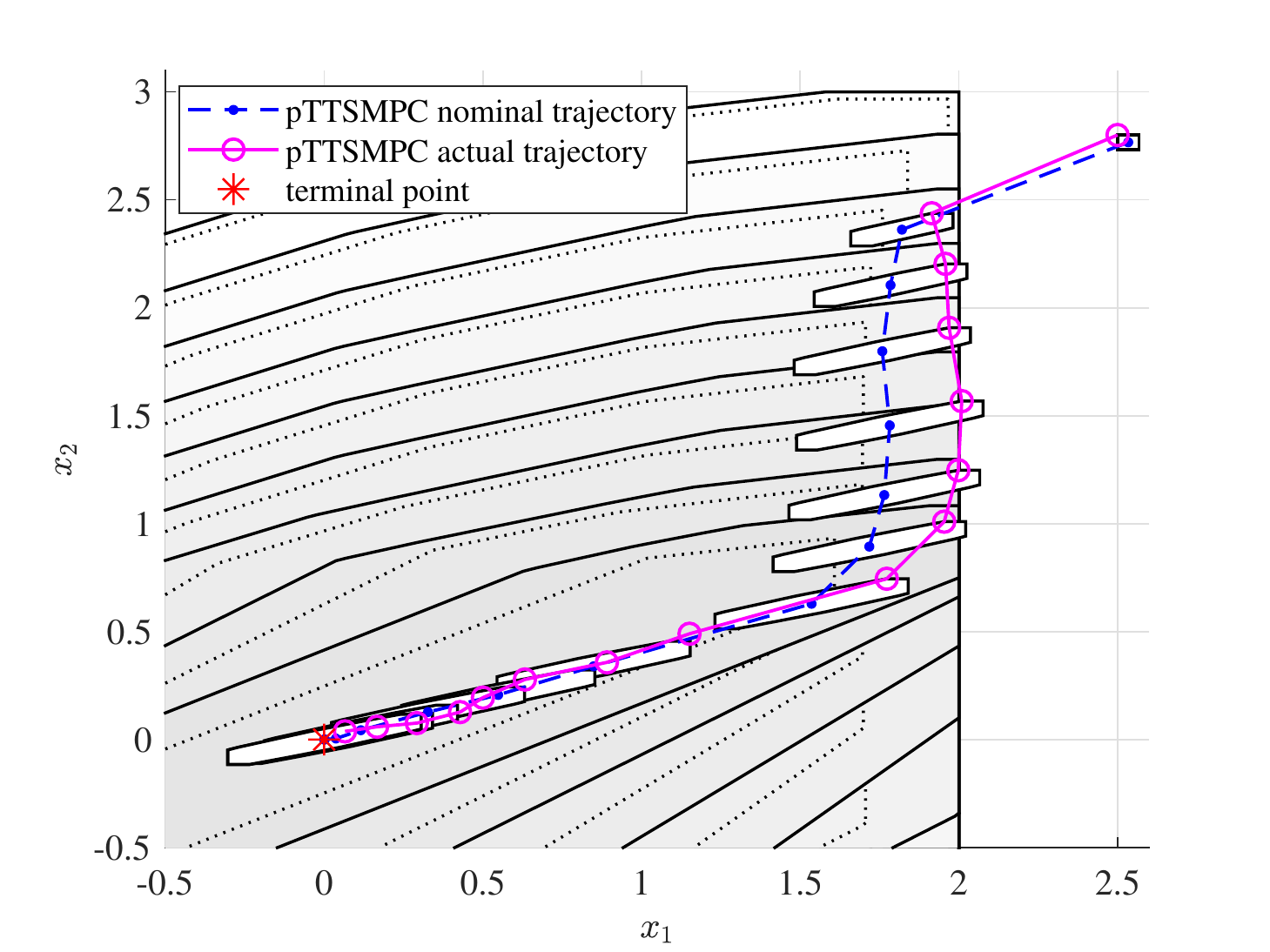}
\end{minipage}
  \caption{Closed-loop state trajectories of the pTTSMPC-en.}\label{Fig trajectory Vs}
\end{figure}

\begin{figure}[tbhp]
\begin{minipage}{1\linewidth}
 \centering
  \includegraphics[width=0.95\textwidth,trim= 0 0 0 0,clip]{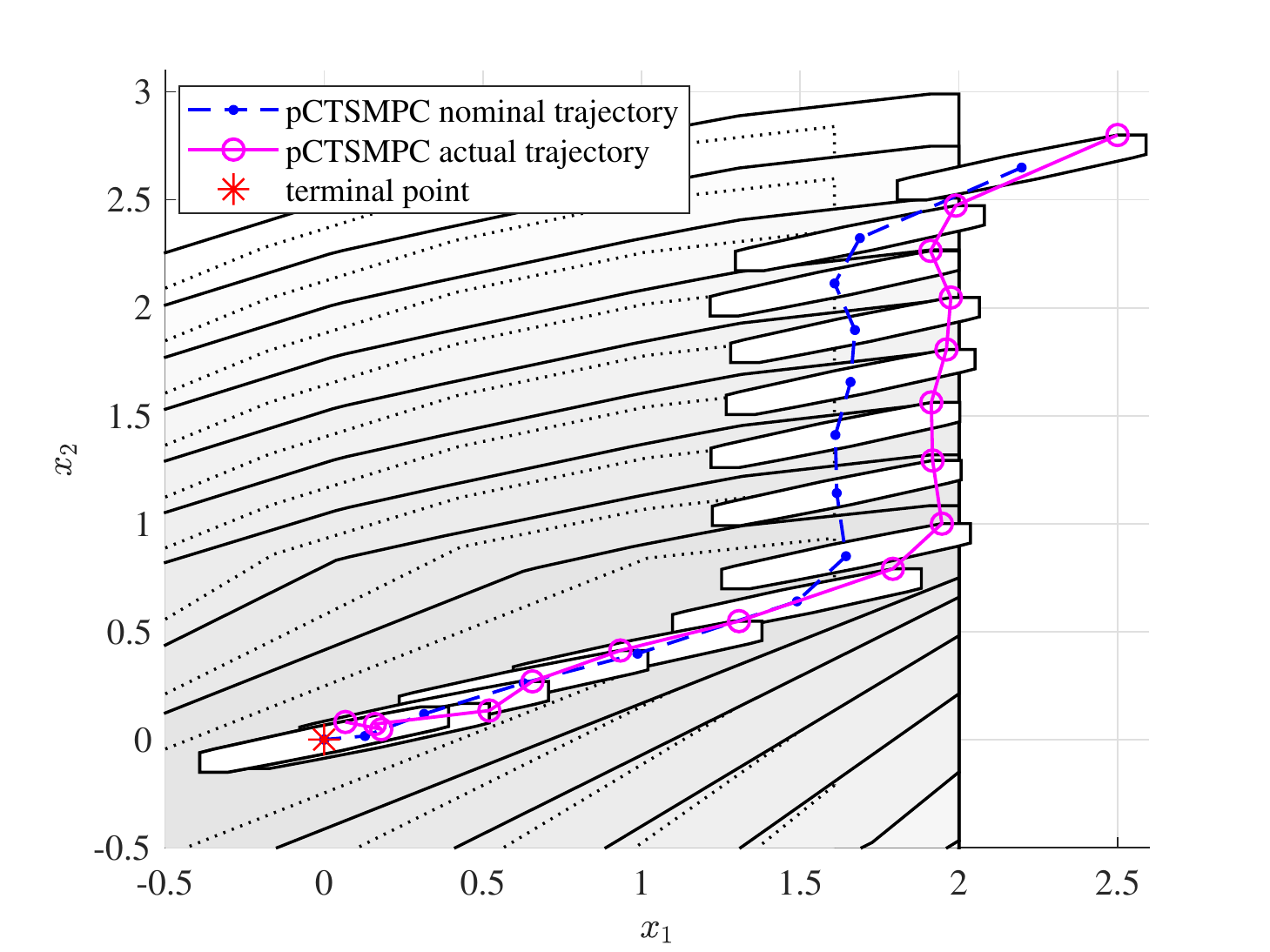}
\end{minipage}
  \caption{Closed-loop state trajectories of the pCTSMPC.}\label{Fig trajectory Cp}
\end{figure}

\begin{figure}[tbhp]
\begin{minipage}{1\linewidth}
 \centering
  \includegraphics[width=0.95\textwidth,trim= 0 0 0 0,clip]{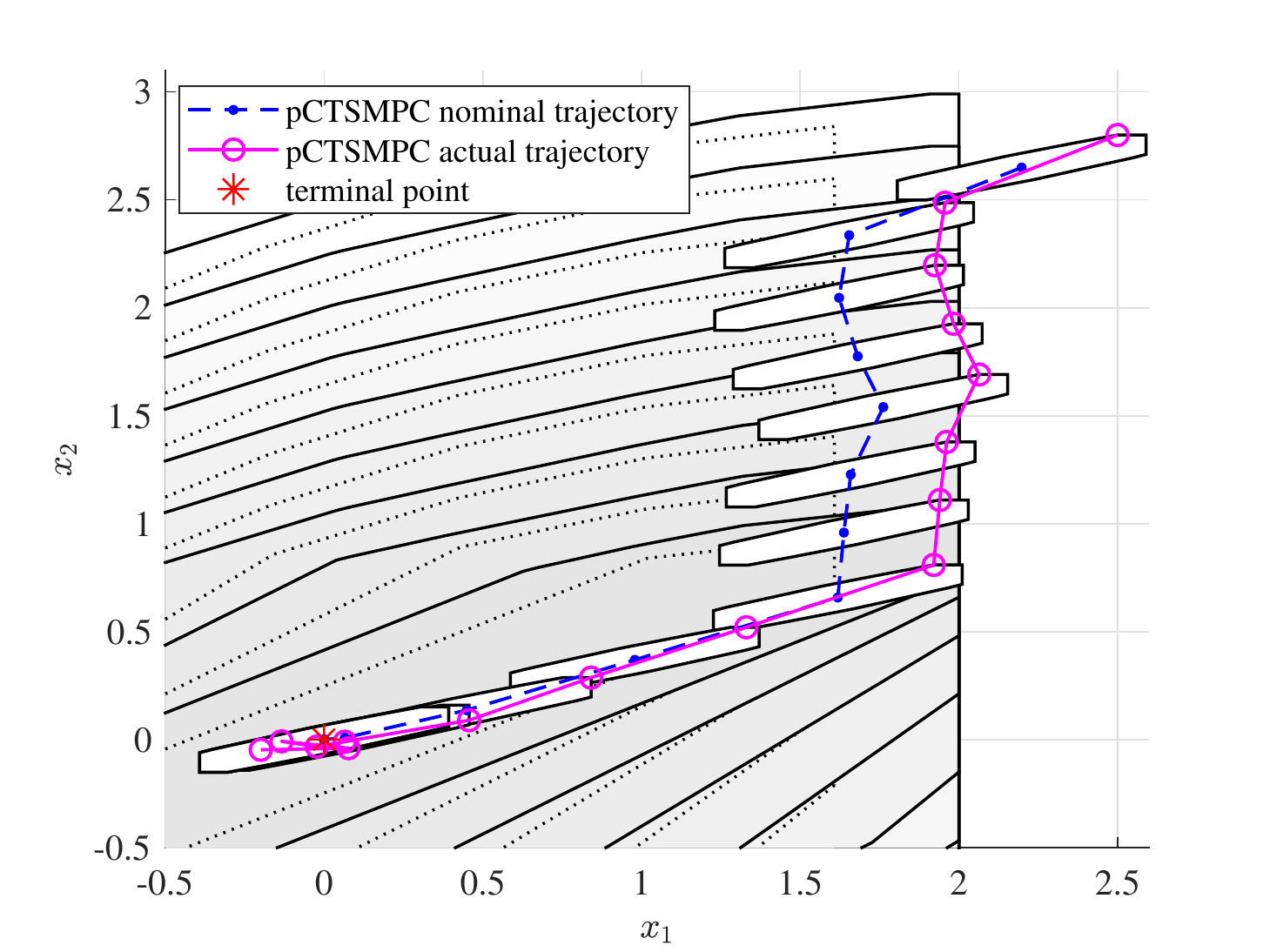}
\end{minipage}
  \caption{Closed-loop state trajectories of the pCTSMPC-en.}\label{Fig trajectory Cs}
\end{figure}

Figs. \ref{Fig trajectory Vp}-\ref{Fig trajectory Cs} present the closed-loop state trajectories and the relevant initial tubes at time instant $ t $, $ t \in \mathbb N_0^{N_{\rm sim}-1} $, of the four probabilistic tube-based SMPC schemes, respectively. In the figures, the solid-circle line (red) represents the actual trajectories, while the dash-dot line (blue) is the nominal trajectories. The initial tubes $ s_{0|t} \oplus \mathcal E^{e_t}_{1-\varepsilon} $ and $ s_{0|t} \oplus \lambda \mathcal E^{e_t}_{1-\varepsilon} $ at time instant $ t $ are shown as the small hexagons along the nominal trajectories in Fig. \ref{Fig trajectory Vp} and Fig. \ref{Fig trajectory Vs}. It can be seen that the tubes are varying over time to guarantee the feasibility. And the tube of the pTTSMPC-en approach is almost the same as that of the pTTSMPC version at the same time instant, which means the slack variable $ \lambda $ is enforced to tend towards 1 at all the simulation time instants.

Analogously, the initial tubes $ s_{0|t} \oplus Z $ and $ s_{0|t} \oplus \lambda Z $ are shown as the small hexagons along the nominal trajectories in Fig. \ref{Fig trajectory Cp} and Fig. \ref{Fig trajectory Cs}, respectively. Here, the tubes of the pCTSMPC approach are constant over time. And the relevant time instant tubes of the pCTSMPC-en approach are almost the same as that of the pCTSMPC version, which means the slack variable $ \lambda $ is enforced to tend towards 1 at all the simulation time instants.

Furthermore, in order to confirm the feasibility of the proposed probabilistic tube-based SMPC schemes, we conduct $ N_{\rm s}=10000 $ simulations based on the same parameter setup. The feasibility ratio is defined as the comparative value of the number of feasibility $ n_f $ to the total simulation number $ N_{\rm s} $, i.e.,
\begin{equation*}
r_f = \frac{n_f}{N_{\rm s}} \times 100\%.
\end{equation*}
Table \ref{table_1} illustrates the feasibility ratio of the four approaches. Unexpectedly, the experimental feasibility ratios of all the probabilistic tube-based SMPC algorithms for the system in (\ref{example_system_1}) are 100\%, that is to say all the schemes possess good practical feasibility. 

\begin{table}[tbhp]
\renewcommand{\arraystretch}{1.3}
\caption{Feasibility Ratio of Probabilistic Tube-based SMPC Algorithms with 10000 Realizations}\label{table_1}
\centering
\setlength{\tabcolsep}{1mm}
{
\begin{tabular}[]{c c c c c}
\toprule
  algorithm   & pTTSMPC &  pTTSMPC-en  & pCTSMPC &  pCTSMPC-en \\
\midrule
  $ r_f $       & 100\%     &      100\%     &     100\%   &     100\%    \\
\bottomrule
\end{tabular}
}
\end{table}

\subsection{Constraint Violation}

To verify the chance constraint satisfaction of the proposed probabilistic tube-based SMPC schemes, we introduce three definitions, namely, the empirical maximum constraint violation ratio, the empirical minimum constraint violation ratio and the empirical average constraint violation ratio. The number of violation at time instant $ t $ over $ N_{\rm s} $ simulations is denoted as $ n_v(t) $. Then the empirical maximum constraint violation ratio is defined as $  r^{max}_v \triangleq \max r_v(t) $, $ t \in \mathbb N_0^{N_{\rm sim}-1} $, with the constraint violation ratio at time instant $ t $
\begin{equation*}
r_v(t) = \frac{n_v(t)}{N_{\rm s}} \times 100\%.
\end{equation*}
The empirical minimum constraint violation ratio is defined as $  r^{min}_v \triangleq \min r_v(t) $, $ t \in \mathbb N_1^6 $.
And the empirical average constraint violation ratio $ \bar r_v  $ of the proposed algorithms for the system in (\ref{example_system_1}) are calculated as the mean of the constraint violation ratios at the first 6 time instants.

Table \ref{table_2} illustrates the empirical average constraint violation ratio $ \bar r_v  $, the empirical maximum constraint violation ratio $  r^{max}_v  $, the empirical minimum constraint violation ratio $ r^{min}_v $ and the total time consumption of the probabilistic tube-based SMPC algorithms for $ N_{\rm s}=10000 $ simulations with the same parameter setup. From the table we can see that:
1) All the proposed probabilistic tube-based SMPC algorithms can guarantee the chance constraint satisfaction.
2) The ratios $ \bar r_v  $, $ r^{max}_v $, $ r^{min}_v $ of the strongly feasible tube-based SMPC algorithms are smaller than those of the related probabilistically resolvable versions due to the slackness of the state initializations.
3) Because of the conservatism of the constant tightening, the constant tube-based SMPC approaches give smaller $ \bar r_v  $, $ r^{max}_v $, $ r^{min}_v $ than the corresponding time-varying tube-based SMPC versions.
4) In spite of resulting in more conservatism, the constant tube-based SMPC approaches require less time consumptions than the corresponding time-varying tube-based SMPC approaches.

\begin{table}[tbhp]
\renewcommand{\arraystretch}{1.3}
\caption{Constraint Violation and Time Consumption of Probabilistic Tube-based SMPC Algorithms with 10000 Realizations}\label{table_2}
\centering
\setlength{\tabcolsep}{1mm}
{
\begin{tabular}[]{c c c c c}
\toprule
  algorithm   & pTTSMPC &  pTTSMPC-en  & pCTSMPC &  pCTSMPC-en \\
\midrule
  $ \bar r_v $   & 19.71\%    &   19.58\%      & 18.09\%     &  17.97\%    \\
  $ r^{max}_v $  & 20.88\%    &   20.45\%      & 18.42\%     &  18.08\%    \\
  $ r^{min}_v $  & 18.66\%    &   18.62\%      & 17.92\%     &  17.77\%    \\  
   time(s)       & 17042      &   18453        & 14853       &  13449      \\  
\bottomrule
\end{tabular}
}
\end{table}

\begin{figure}[tbhp]
\begin{minipage}{1\linewidth}
 \centering
  \includegraphics[width=0.95\textwidth,trim= 0 0 0 0,clip]{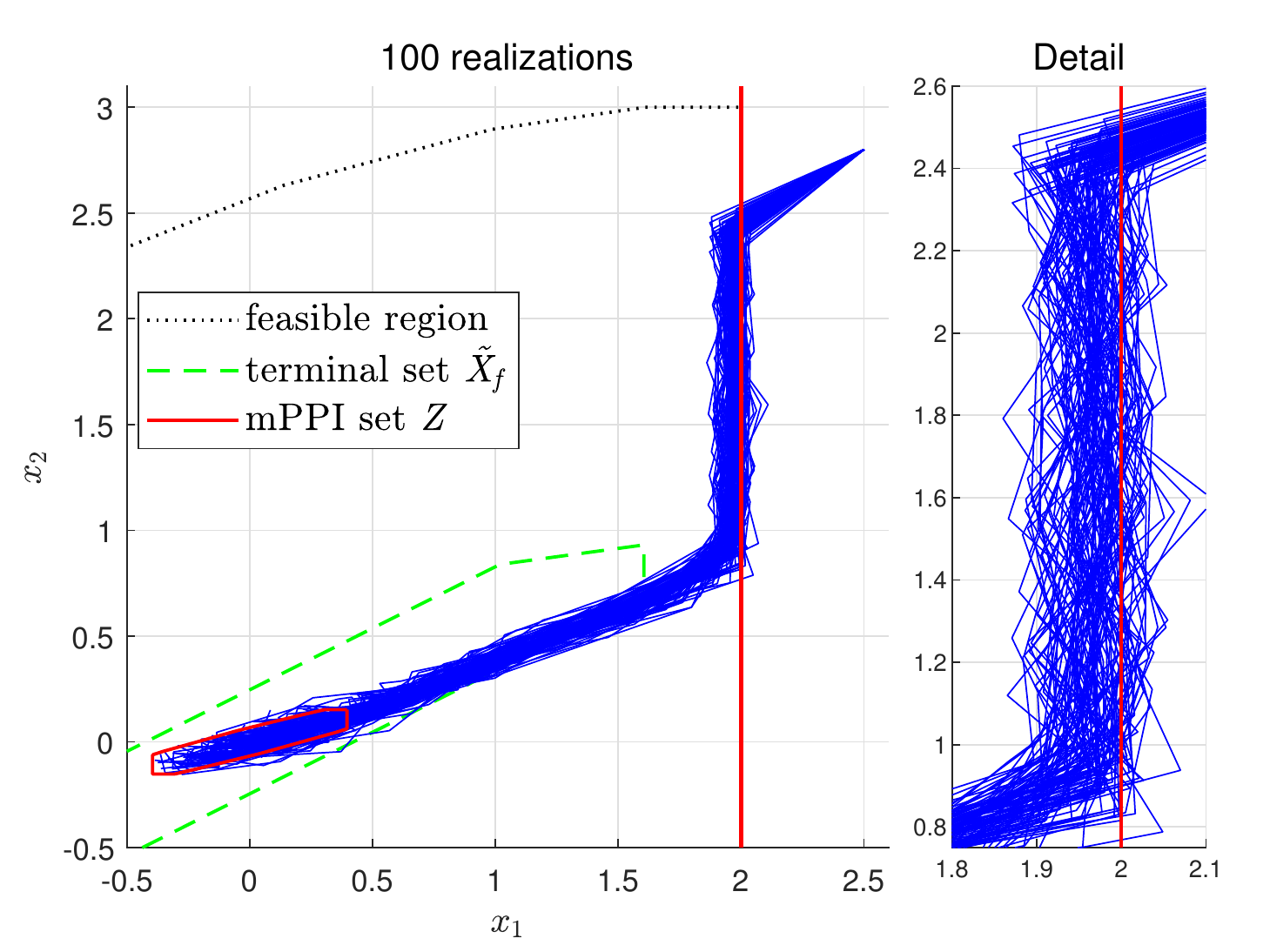}
\end{minipage}
  \caption{Chance constraint violation of pTTSMPC.}\label{Fig violation Vp}
\end{figure}

\begin{figure}[tbhp]
\begin{minipage}{1\linewidth}
 \centering
  \includegraphics[width=0.95\textwidth,trim= 0 0 0 0,clip]{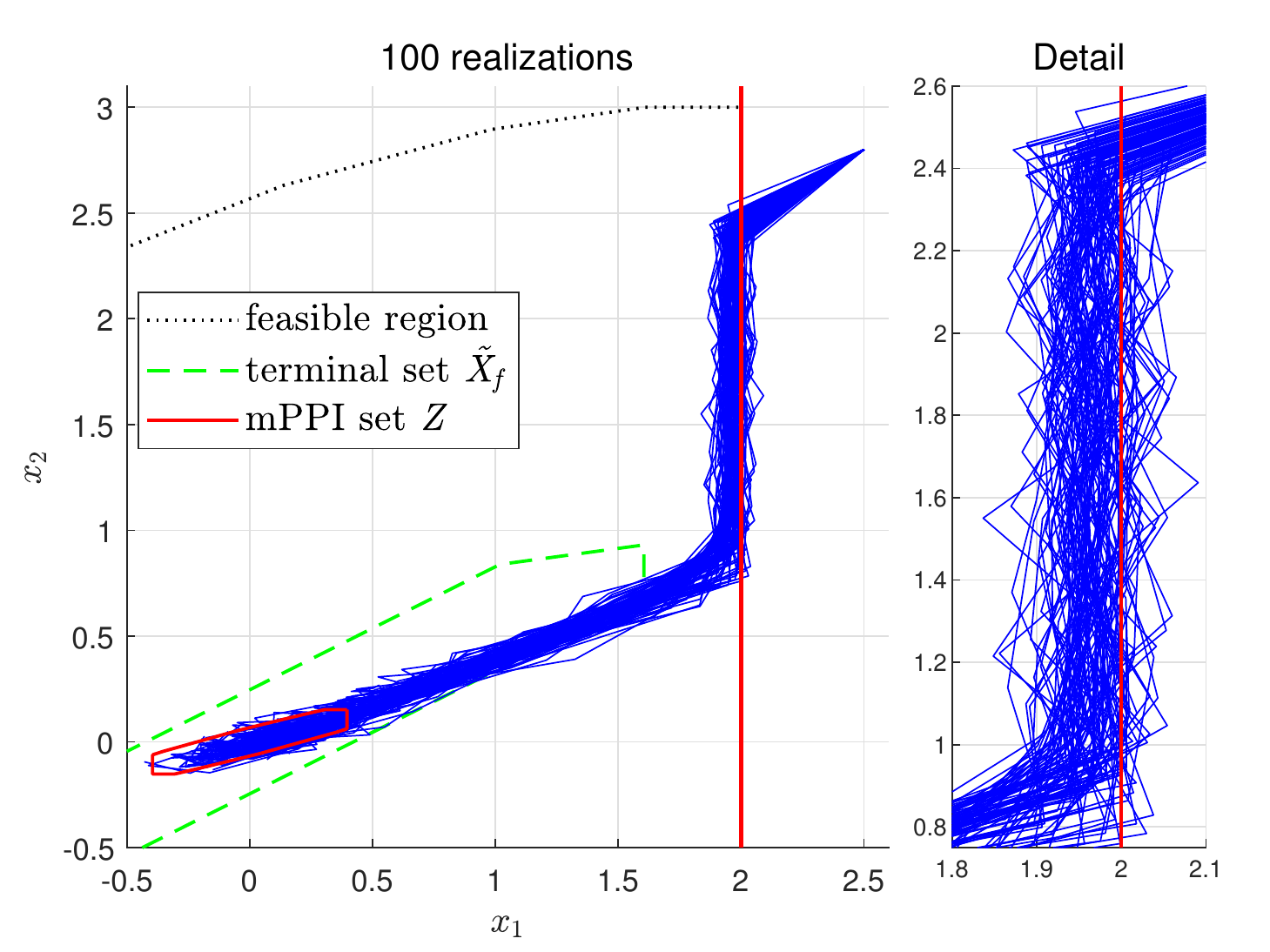}
\end{minipage}
  \caption{Chance constraint violation of pTTSMPC-en.}\label{Fig violation Vs}
\end{figure}

\begin{figure}[tbhp]
\begin{minipage}{1\linewidth}
 \centering
  \includegraphics[width=0.95\textwidth,trim= 0 0 0 0,clip]{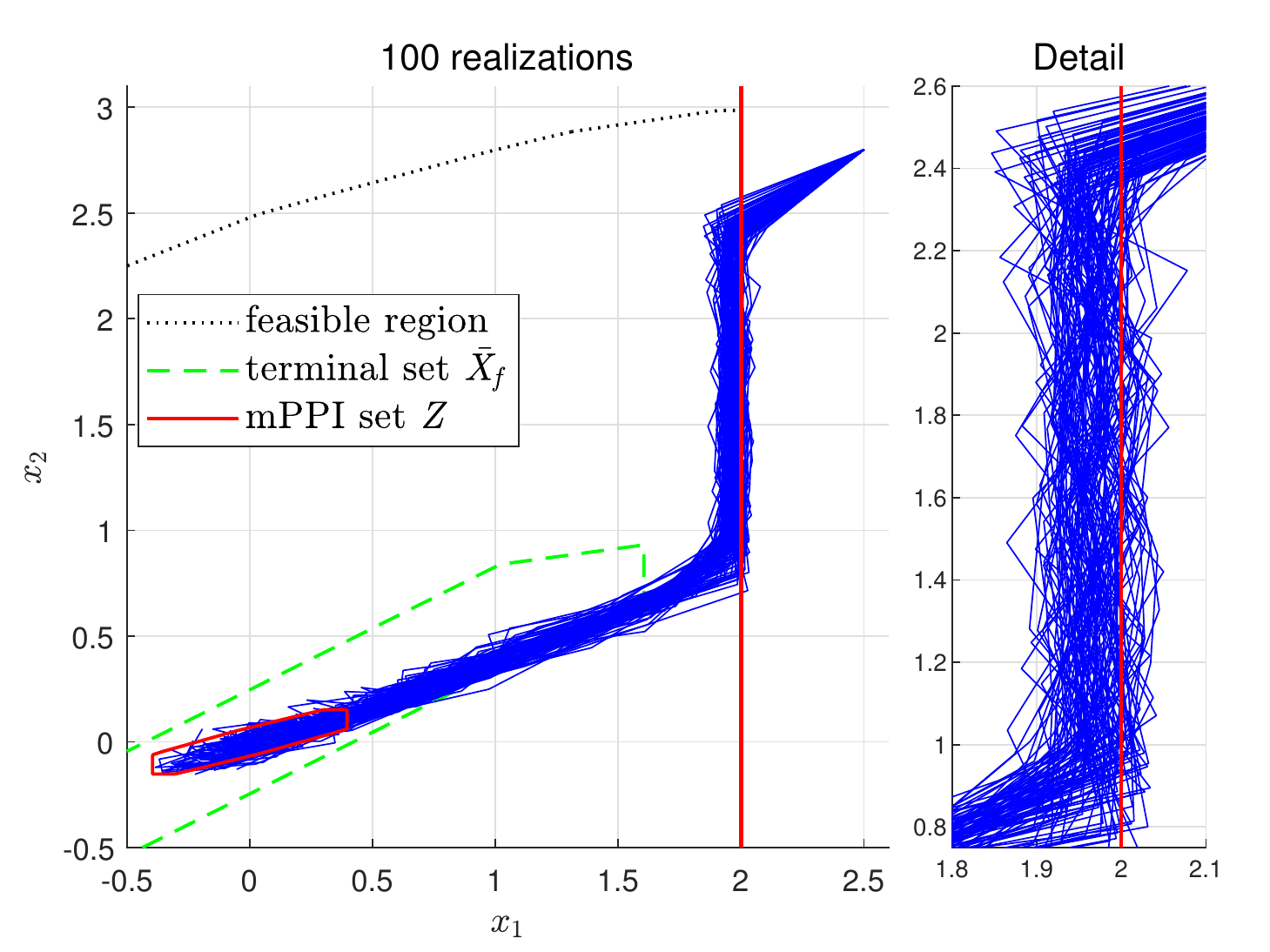}
\end{minipage}
  \caption{Chance constraint violation of pCTSMPC.}\label{Fig violation Cp}
\end{figure}

\begin{figure}[tbhp]
\begin{minipage}{1\linewidth}
 \centering
  \includegraphics[width=0.95\textwidth,trim= 0 0 0 0,clip]{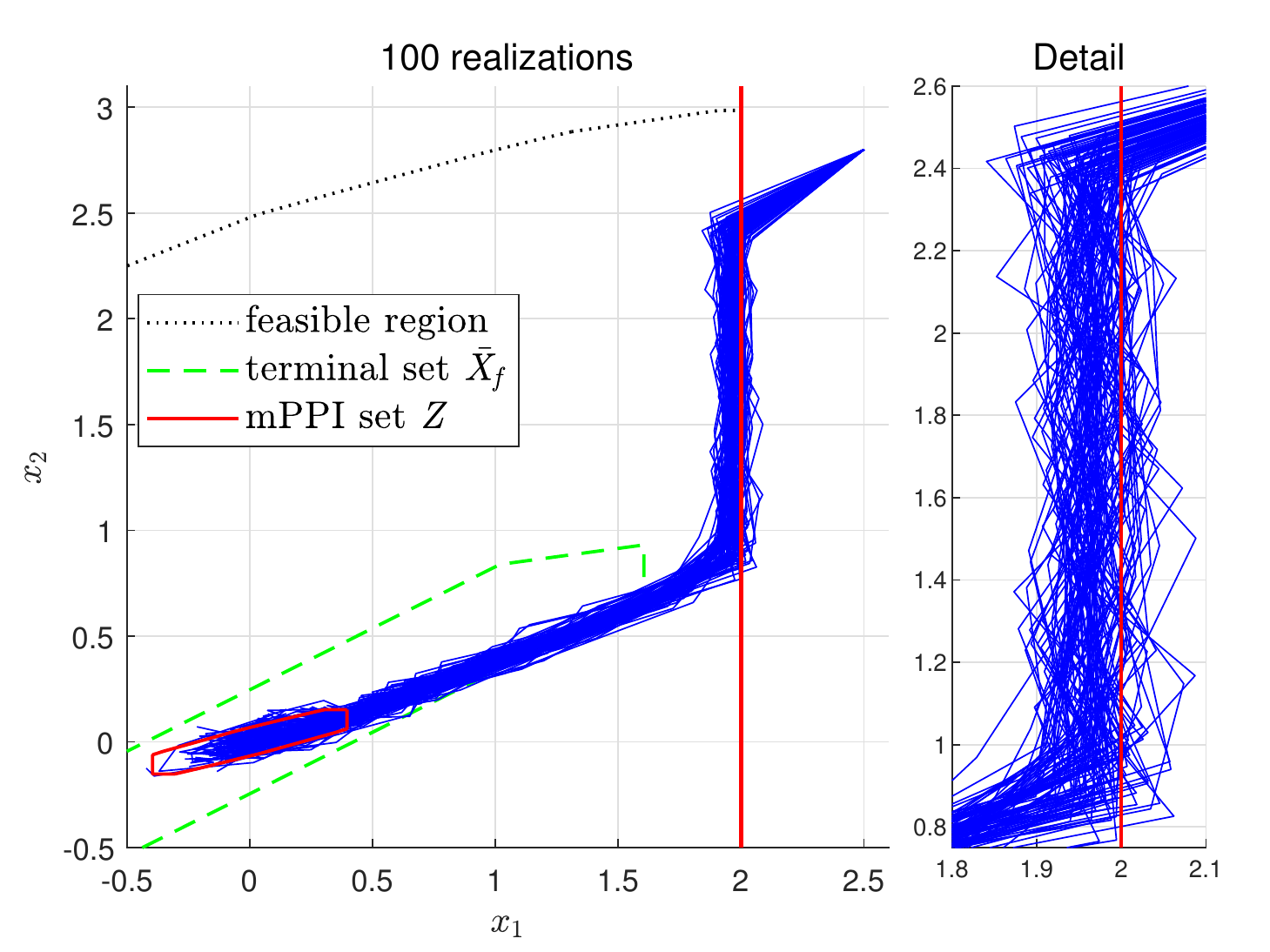}
\end{minipage}
  \caption{Chance constraint violation of pCTSMPC-en.}\label{Fig violation Cs}
\end{figure}

To clearly illustrate the constraint violation of the four algorithms, Figs. \ref{Fig violation Vp}-\ref{Fig violation Cs} report the simulation results of them with 100 realizations, respectively. For each figure, the left part demonstrates the closed-loop trajectories with 15-step simulation, while the right part exhibits the detailed constraint violation. It can be seen that, in addition to the satisfaction of chance constraint, the closed-loop trajectories of the actual state will stay in the mPPI set ultimately with a probability greater than the prescribed 80\%.

\subsection{Comparison of pTTSMPC Schemes with Different Initialization Methods}

To verify the advantage of the proposed flexible initialization method over the existing ones, we further compare the control performance of pTTSMPC schemes with different initialization methods. That is to say, the optimal control problem in (\ref{tTSMPC}) with different initial state constraints will be demonstrated for the system in (\ref{example_system_1}) under the same parameter setup. Before that, we define four initialization methods as follows.
\begin{itemize}[leftmargin=13mm]
	\item[ Case1: ] The proposed flexible initialization in (\ref{constraints init v}), namely $  x_t - s_{0|t} \in \mathcal D_t $. 
	\item[ Case2: ] The initial nominal state is set to be the predicted value at the previous time instant, namely $ s_{0|t} = s_{1|t-1} $.
	\item[ Case3: ] The initialization for the nominal state is set as the predicted value at the first time instant as in \cite{Hewing2020}, i.e., $ s_{0|t} = s_{t|0} $.
	\item[ Case4: ] The initialization for the nominal state is chosen from the alternative strategies as in \cite{Farina2013, Farina2015, Hewing2018b}. That is, $ s_{0|t} = x_t $ whenever feasible and $ s_{0|t} = s_{1|t-1} $ otherwise.
\end{itemize}

Table \ref{table_3} illustrates the empirical average constraint violation ratio $ \bar r_v  $, the empirical maximum constraint violation ratio $  r^{max}_v  $, the empirical minimum constraint violation ratio $ r^{min}_v $ and the total time consumption of the four pTTSMPC algorithms with different initialization methods based on $ N_{\rm s}=10000 $ simulations. From the table we can see that:
1) Case1 gives the largest empirical constraint violation ratios, which is less conservative than other cases. Meanwhile, the time consumption of Case1 is heavier than those of other cases.
2) For the other three cases, the computational loads are comparable. The violations of Case2 and Case3 are obviously lower than those of Case1 due to the lack of feedback information from the recently measured state. Case4 is the most conservative one in comparison with others and suffers possible infeasibility owing to the fully feedback.

\begin{table}[tbhp]
\renewcommand{\arraystretch}{1.3}
\caption{Performance of pTTSMPC Schemes with Different Initialization Methods - 10000 Realizations}\label{table_3}
\centering
\setlength{\tabcolsep}{1mm}
{
\begin{tabular}[]{c c c c c}
\toprule
  algorithm      & Case1       &  Case2       & Case3      &  Case4 \\
\midrule
  $ \bar r_v $   &  19.71\%    & 12.70\%      &  6.90\%    &  0.007\%    \\
  $ r^{max}_v $  &  20.88\%    & 16.32\%      & 12.09\%    &   0.02\%    \\
  $ r^{min}_v $  &  18.66\%    &     0\%      &     0\%    &      0\%    \\  
   time(s)       &  17042      &  13670       &  13677     &  14041      \\  
\bottomrule
\end{tabular}
}
\end{table}

\section{Conclusion} \label{Conclusion}
Utilizing the concept of PRS, two probabilistically resolvable tube-based stochastic MPC approaches have been developed for linear constrained system with additive unbounded stochastic uncertainties. The results on the feasibility and closed-loop stability are presented. In comparison with existing result, the proposed method can reduce the conservatism in chance constraints satisfaction. The simulation studies verified the feasibility and advantage of the proposed methods. The future work will be the stochastic MPC for nonlinear systems with unbounded disturbances.

%% use section* for acknowledgment
%\section*{Acknowledgment}
%
%This work was supported in part by the National Natural Science Foundation of China (NSFC) under Grant 61922068, 61733014; in part by Shaanxi Provincial Funds for Distinguished Young Scientists under Grant 2019JC-14; in part by Aoxiang Youth Scholar Program under Grant 20GH0201111.

% Can use something like this to put references on a page
% by themselves when using endfloat and the captionsoff option.
\ifCLASSOPTIONcaptionsoff
  \newpage
\fi

% trigger a \newpage just before the given reference
% number - used to balance the columns on the last page
% adjust value as needed - may need to be readjusted if
% the document is modified later
%\IEEEtriggeratref{8}
% The "triggered" command can be changed if desired:
%\IEEEtriggercmd{\enlargethispage{-5in}}

%%%%%%%%%%%%%%%%%%%%%%%%%%%%%%%%%%%%%%%%%%%%%%%%%%%%%%%%%%%%%%%%%%%
% references section

% biography section
% 
% If you have an EPS/PDF photo (graphicx package needed) extra braces are
% needed around the contents of the optional argument to biography to prevent
% the LaTeX parser from getting confused when it sees the complicated
% \includegraphics command within an optional argument. (You could create
% your own custom macro containing the \includegraphics command to make things
% simpler here.)
%\begin{IEEEbiography}[{\includegraphics[width=1in,height=1.25in,clip,keepaspectratio]{mshell}}]{Michael Shell}

%\enlargethispage{-5in}

% that's all folks
\end{document}